\documentclass[pra,reprint,showkeys,showpacs,times,superscriptaddress]{revtex4-1}
\usepackage{amsmath}
\usepackage{amssymb}
\usepackage{amsthm}
\usepackage{times}
\usepackage{graphics}
\usepackage{graphicx}
\usepackage{bm}
\usepackage{subfigure}
\usepackage{enumerate}
\usepackage[pagebackref=false]{hyperref} 
\hypersetup{
    bookmarks=true,         
    unicode=false,          
    pdftoolbar=true,        
    pdfmenubar=true,        
    pdffitwindow=false,     
    pdfstartview={FitH},    
    pdfkeywords={keyword1} {key2} {key3}, 
    pdfnewwindow=true,      
    colorlinks=true,       
    linkcolor=red,          
    citecolor=blue,        
    filecolor=magenta,      
    urlcolor=cyan,           
}

\def\ket#1{ | #1 \rangle}
\def\bra#1{{\langle #1 |  }}

\def\pd2v#1#2#3{\frac{\partial^2 #1}{\partial #2 \partial #3}}

\def \2x2mat#1#2#3#4{
\left( \begin{array}{cc}
#1 &  #2 \\  #3 &  #4
\end{array} \right)
}

\def \i{i}

\usepackage{lipsum}
\usepackage{xcolor}
\usepackage{xspace}

\newcommand{\red}{\textcolor{black}}

\begin{document}


\title{Visualising multiqubit correlatons using the Wigner function.}
\author{Todd Tilma}
\affiliation{Tokyo Institute of Technology, 2-12-1 Ookayama, Meguro-ku, Tokyo 152-8550, Japan}
\affiliation{Quantum Systems Engineering Research Group \& Department of Physics, Loughborough University, Leicestershire LE11 3TU, United Kingdom}
\author{Mario A. Ciampini}
\affiliation{Vienna Center for Quantum Science and Technology (VCQ), Faculty of Physics, University of Vienna, Boltzmanngasse 5, A-1090 Vienna, Austria}
\affiliation{Dipartimento di Fisica, Sapienza Universit\`a di Roma, P.le Aldo Moro 5, 00185, Rome, Italy}
\author{Mark J. Everitt}
\affiliation{Quantum Systems Engineering Research Group \& Department of Physics, Loughborough University, Leicestershire LE11 3TU, United Kingdom}
\author{W. J. Munro}
\affiliation{NTT Basic Research Labs \& NTT Research Center for Theoretical Quantum Physics, NTT Corporation, 3-1 Morinosato-Wakamiya, Atsugi, Kanagawa 243-0198, Japan}
\affiliation{National Institute of Informatics, 2-1-2 Hitotsubashi, Chiyoda-ku, Tokyo 101-8430, Japan}
\author{Paolo Mataloni}
\affiliation{Dipartimento di Fisica, Sapienza Universit\`a di Roma, P.le Aldo Moro 5, 00185, Rome, Italy}
\author{Kae Nemoto}
\affiliation{National Institute of Informatics, 2-1-2 Hitotsubashi, Chiyoda-ku, Tokyo 101-8430, Japan}
\author{Marco Barbieri}
\affiliation{Dipartimento di Scienze, Universit\`a degli Studi Roma Tre, Via della Vasca Navale 84, 00146, Rome, Italy}

\date{\today}

\begin{abstract}
\red{Quantum engineering now allows  to design and construct multi-qubit states in a range of physical systems. These states are typically quite complex in nature, with disparate, but relevant properties that include  both single and multi-qubit coherences and even entanglement. All these properties can be assessed by reconstructing the density matrix of those states -- but the large parameter space can mean physical insight of the nature of those states and their coherence can be hard to achieve.} Here we explore how the Wigner function of a multipartite system and its visualization provides rich information on the \red{nature of the state, not only at} illustrative level \red{but also at the quantitative level. We test our tools in} a photonic architecture making use of the multiple degrees of freedom of two photons.
\end{abstract}


\maketitle

\section{Introduction}\label{Intro}

\red{We have now entered a quantum technology era~\cite{Dowling1655}, in which the ability to create, manipulate, \red{measure} and characterize quantum states and processes is critical to improve performance in cryptography, communication, computing, and metrology. For research purposes, the reliability of a quantum device is often assessed by reconstructing either its state or the process it performs by means of quantum tomography~\cite{Breitenbach97,White99,Banaszek00,James01,Dariano03,ParisBook,Blume10,Rehacek10,Cooper14}. This technique employs a series of measurements on a large enough number of identically prepared copies of the quantum system in order to derive the matrix associated to either the quantum state or process. This knowledge is key to access in an experiment all those quantities that can not be directly measured, foremost the type and amount of entanglement \cite{Vedral97,Wootters98,Dur00,Giedke01}, but this extends to the entropy~\cite{NielsonChuang}, and to different signatures of nonclassicality~\cite{PhysRev.131.2766,PhysRevLett.10.277,Vogel00,Ourjoumtsev06,Haroche08}. For small systems and simple processes, the density and process matrices also provides some visual help when it comes to inspecting and comparing them to some reference. This ability, however, is not easily transposed as the size of the quantum object grows; in addition, the complex nature of the off diagonal elements of the matrices makes their interpretation even harder them.} 

\red{As density and process matrices have been introduced as tools for calculations, they do not necessarily represent the best means to visualising quantum states. Even for simple systems such as qubits, processes are more easily captured by their Bloch sphere picture. In the quantum optics community, instead, there have been a number of phase space approaches that lend themselves naturally to a pictorial representations. These include the Glauber-Sudarshan $P$, Husimi $Q$, and Wigner representations, each enjoying a number of interesting attributes~\cite{PerelomovB}. They have been used extensively in the continuous variable (CV) formalism~\cite{Braunstein05,Weedbrook12}, for which the dimensionality of the Hilbert space makes the density matrix approach very limited. In particular, the Wigner function~\cite{Wigner1932} has been long been employed as the preferred mean for the phase space representation of quantum states in optical and microwave fields~\cite{Lvovsky09,Wallraff,Clealand}. The preferential choice for this particular quasi-probability distribution stems from two main reasons: the marginal distributions are genuine probability distributions giving information about the state being examined~\cite{Wigner1932}, and, more intriguingly, the presence of regions where the Wigner function attains negative values is a valid indication of the so called  {\it quantumness} of the state~\cite{Ourjoumtsev06}, since, in this case, negative regions cannot occur within the real probability distribution of our classical world.}

\red{Given the usefulness of this approach in the CV world, it would thus be natural to consider whether this carries across into discrete variable systems. The extension of the Wigner functions  to discrete variable systems based on parity operators~\cite{1601.07772,1605.08922} has recently yielded such a function with all the distinctive features seen in the continuous variable regime. } The number of arguments of the Wigner function grows with the number of subsystems, and one has to consider a function of four variables even for two qubits. The informative content of the Wigner function would be lost unless one finds a convenient, compact representation.

\red{In this article we discuss how a compact representation can be obtained by using ``equal-angle-slicing", able to to distinguish distinct classes of quantum states.}  We apply our method to the analysis of three-qubit \red{GHZ and W} entangled states~\cite{PhysRevA.62.062314}) generated in a single-photon based quantum photonics experiment~\cite{Ciampini17}. Our results demonstrate that the Wigner representation is as informative for discrete systems, as it is for continuous ones. We thus expect this to become a commonplace tool, especially in the description of hybrid systems \cite{Bimbard10,Wallraff12,Jeong14,Morin14} comprising both discrete and continuous parts.

\section{Wigner Function for Qubits}\label{Primer}

\red{It is important here to provide a brief description of our Wigner function - especially as it is not defined based on the canonical $X \& P$ quadratures, as seen in most text books. Instead we will consider the displaced parity version \cite{1601.07772,1605.08922,PhysRevA.15.449,PhysRevA.60.674,PhysRevLett.82.2009,PhysRevLett.105.253603,PhysRevLett.116.133601} :}
\begin{equation}
\label{Wdiff}
W(\Omega)=\frac{1}{(\pi)^n}\text{Tr}\left[\rho D(\Omega)\Pi D^\dag(\Omega)\right]
\end{equation}
\red{where $D(\Omega)$ and $\Pi$ are the displacement and parity operators with $\Omega$ being {\it any} complete parameterisation of the phase space~\cite{PhysRevA.50.4488}. }

\begin{figure}[htb]
\includegraphics[width=0.8\columnwidth, 
]{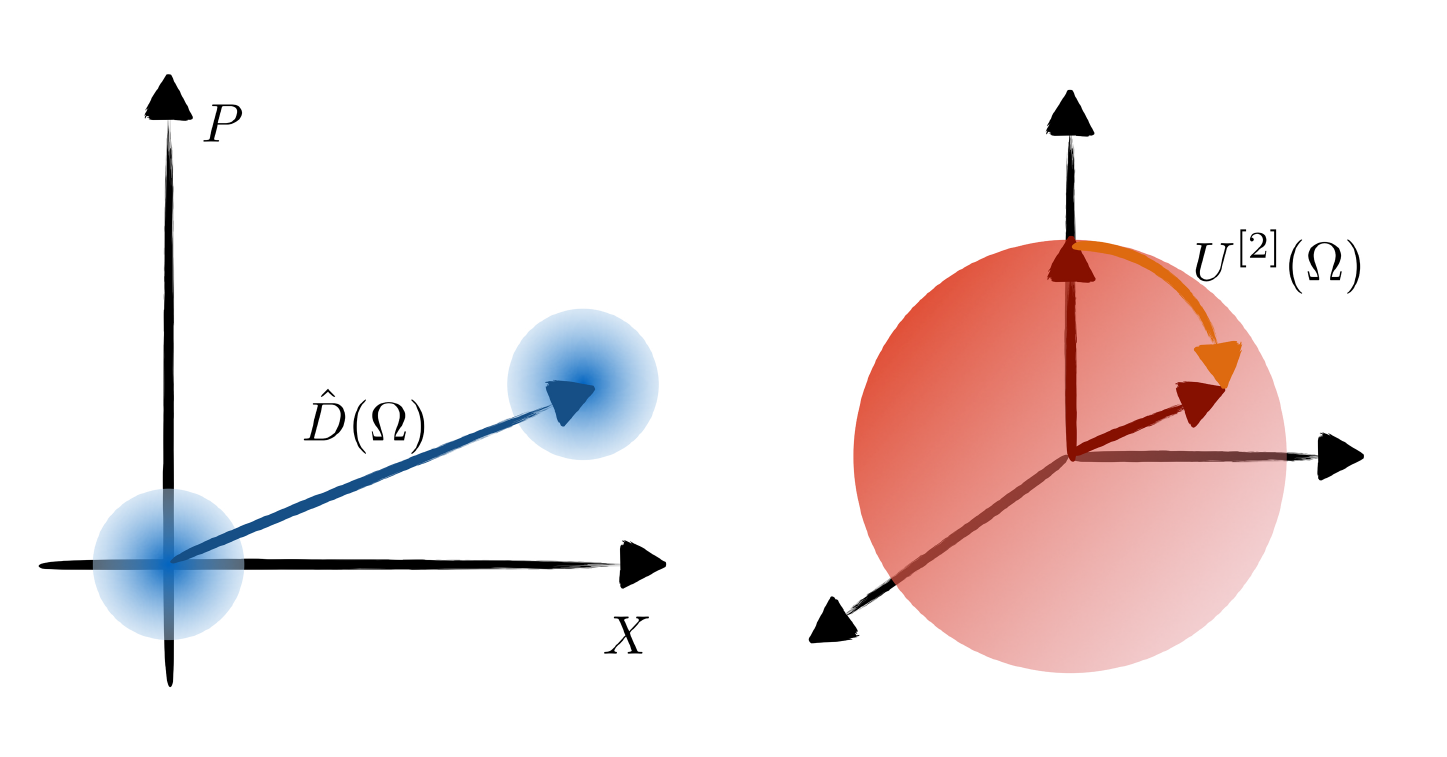}
\caption{Evaluation of the Wigner function. For the standard continuous-variable description (left), the Wigner function can be evaluated as the expectation value of the parity $\Pi$, following a displacement of the state via the operator $D(\Omega)$. For qubits (right), we follow the same strategy: the Wigner function is now defined as the expectation value of the extended parity $\Pi^{[2]}$, following a unitary rotation $U^{[2]}$, playing the role of the displacement on the Bloch sphere.}
\label{figure1}
\end{figure}
\red{These operators  are defined by $D(\Omega)\ket{0}=\ket{\Omega}$ and $\Pi\ket{\Omega}=\ket{-\Omega}$ respectively (see Fig. \ref{figure1}). Now, the Wigner function must display a number of important features:}
\begin{itemize}
\item The representation should be unique;
\item $W(\Omega)$ must be real at all points, and normalised;
\item A rotation on the state $\rho$ must be mapped into a rotation of the phase space parameters;
\item  $\rho$ should be able to be reconstructed from its associated $W(\Omega)$, and vice-versa;
\end{itemize}
\red{These properties are ensured by the appropriate choice of the kernel operator $\Delta(\Omega)\equiv  D(\Omega)\Pi D^\dag(\Omega)/\pi^n$. This indeed plays critical role in the mapping of $\rho$ to $W(\Omega)$ and back. The form of $\Delta(\Omega)$ being composed of displacement and parity operators makes the extension from continuous to discrete variables relatively straightforward. One just needs to define the appropriate $D(\Omega)$ and $\Pi$ in that DV space.} 

\red{It is easier to start with the equivalent displacement operator. In the continuous variable regime the  action of the displacement operator is $D(\Omega)\ket{\Xi}=e^{i \Omega \Xi^\ast} \ket{\Omega+\Xi}$, which effectively translates  from one place in the plane to another. However the set of pure states  for the qubit live on a sphere rather than plane. In such a case  we can} make use of the Euler decomposition of generic single-qubit rotations $U^{[2]}(\theta,\phi,\Phi)=e^{i\sigma_z\phi}e^{i\sigma_y\theta}e^{i\sigma_z\Phi}$ to obtain a parametrization of the phase space.  \red{The appropriate choice of $\theta,\phi,\Phi$ then allows one to rotate from one point on the sphere to another, just like the displacement operation in the CV regime.} 

\red{The second operation we need is a generalized parity operator, analogue to the one in the form $\Pi\ket{\Omega}=\ket{-\Omega}$ for the CV regime.  We have a number of choices available to us, but one that satisfies our criteria above is~\cite{Tilma_2011} $\Pi^{[2]}=\frac{1}{2} \left[ I^{[2]}-\sqrt{3}\sigma_z\right] $ where $\sigma_z$ is the usual Pauli $z$ operator. We can rewrite $\Pi^{[2]}$ as
\begin{eqnarray}
  \Pi^{[2]} = - \left( \begin{array}{cc}  \sqrt{2} \sin(\frac{\pi}{12}) & 0 \\ 0 & \sqrt{2} \cos(\frac{\pi}{12}) \end{array} \right) \sigma_z = {\bf\Theta} \cdot  \sigma_z
\end{eqnarray}
where ${\bf\Theta}$ is the expressed two-dimensional normalization matrix that assists in satisfying the features of our Wigner function based on the given Euler decomposition.
This simplification shows that our \emph{extended parity} effectively imparts a normalized $\pi$-phase shift to spin coherent states.} 

\red{Next, the extension to} $n$ qubits can be obtained simply by using a tensor product~\cite{1605.08922}: 
\begin{equation}
\label{eq:WF1s}
\Delta({\bf \Omega}) \equiv D({\bf \Omega})\Pi D^\dag({\bf \Omega}) = \frac{1}{\upsilon}\bigotimes_{i=1}^{n} D(\Omega_i) \Pi^{[2]} D^\dag(\Omega_i), 
\end{equation}
where $D(\Omega_i) \equiv U^{[2]}(\theta_i,\phi_i,\Phi)$ and $\upsilon$ is a normalization constant dependent on ${\bf \Omega}$. \red{This means we now can calculate the $n$ qubit Wigner function $W(\theta_1,\phi_1, \ldots, \theta_n,\phi_n)$ using (\ref{Wdiff}).}

\section{Slicing the Wigner function}

\red{Now that we have a mechanism to construct an $n$ qubit Wigner function using $n$ pairs of ($\theta$, $\phi$) parameters, it is obvious that} one cannot visualize such large dimensions, \red{thus} it would seem that all the advantage of this pictorial representation would be lost. \red{This is not unique to DV systems, as in the CV regime we are often faced with the problem of illustrating the Wigner function of a multimode system; the adopted solution is to select} opportune ``slices'' over the phase space where such representations are able to provide relevant and valuable information about the system based on physical \red{considerations. A particularly interesting one is the} ``equal-angle'' slice that delivers a visualization depending on only two parameters $\theta$ and $\phi$; \red{this is achieved by setting all $\theta_i = \theta$ and $\phi_i = \phi$}. This slicing is a natural fit to the standard Bloch-sphere representation used in CV, but here the sphere is parametrized by the angles $\theta$ and $\phi$ rather than $X$ and $P$. 
\begin{figure}[htb]
{\includegraphics[width=0.9 \columnwidth, keepaspectratio=true]{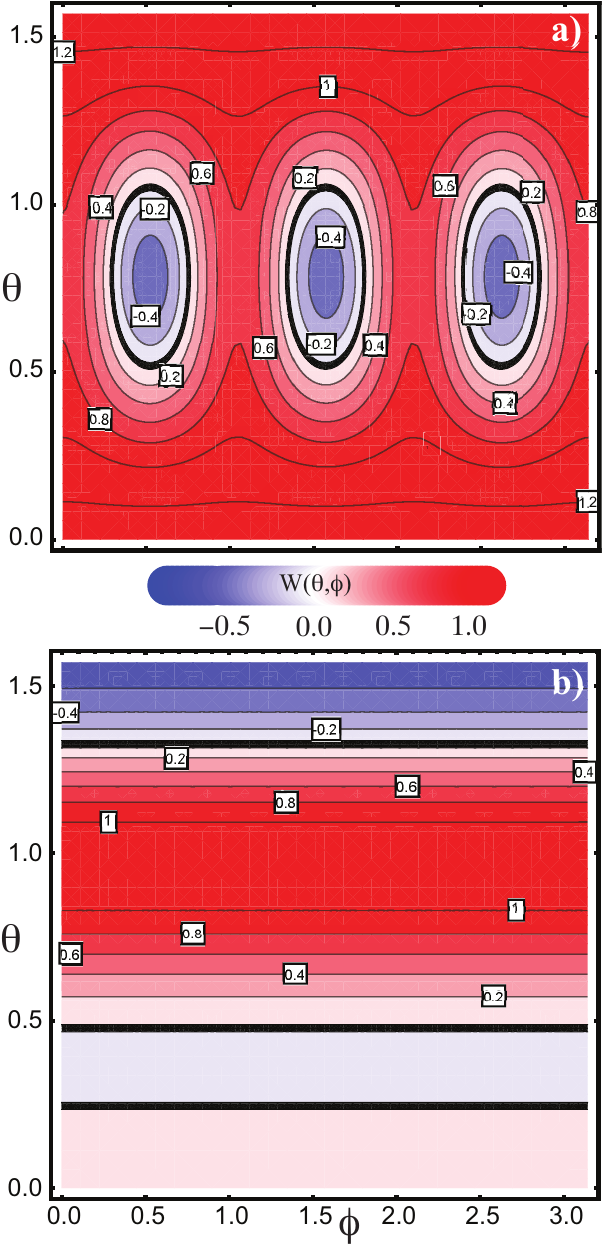}}
\caption{Plot of the magnitude of the ideal equal angle Wigner function of the cluster GHZ (a) and W (b) states versus $\theta$ and $\phi$. Here the red indicates where the Wigner function is positive while blue where it is negative. The black line boundary indicates where the Wigner function is zero.  It is clear to see the difference both in structure and in magnitude of the two considered states, highlighting their distinctive properties at a glance.}
\label{idealwigner}  
\end{figure}
\begin{figure*}[t]
\includegraphics[width=0.9\textwidth]{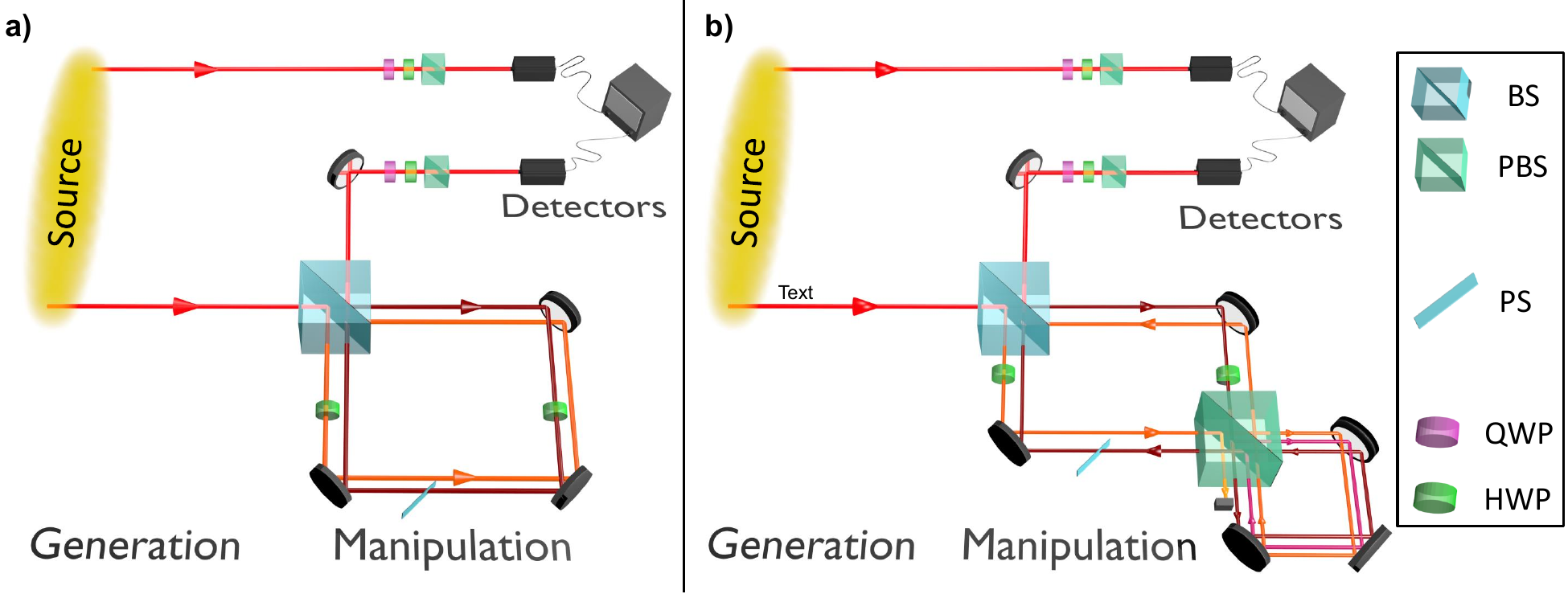}\\
\includegraphics[width=0.9 \columnwidth, keepaspectratio=true]{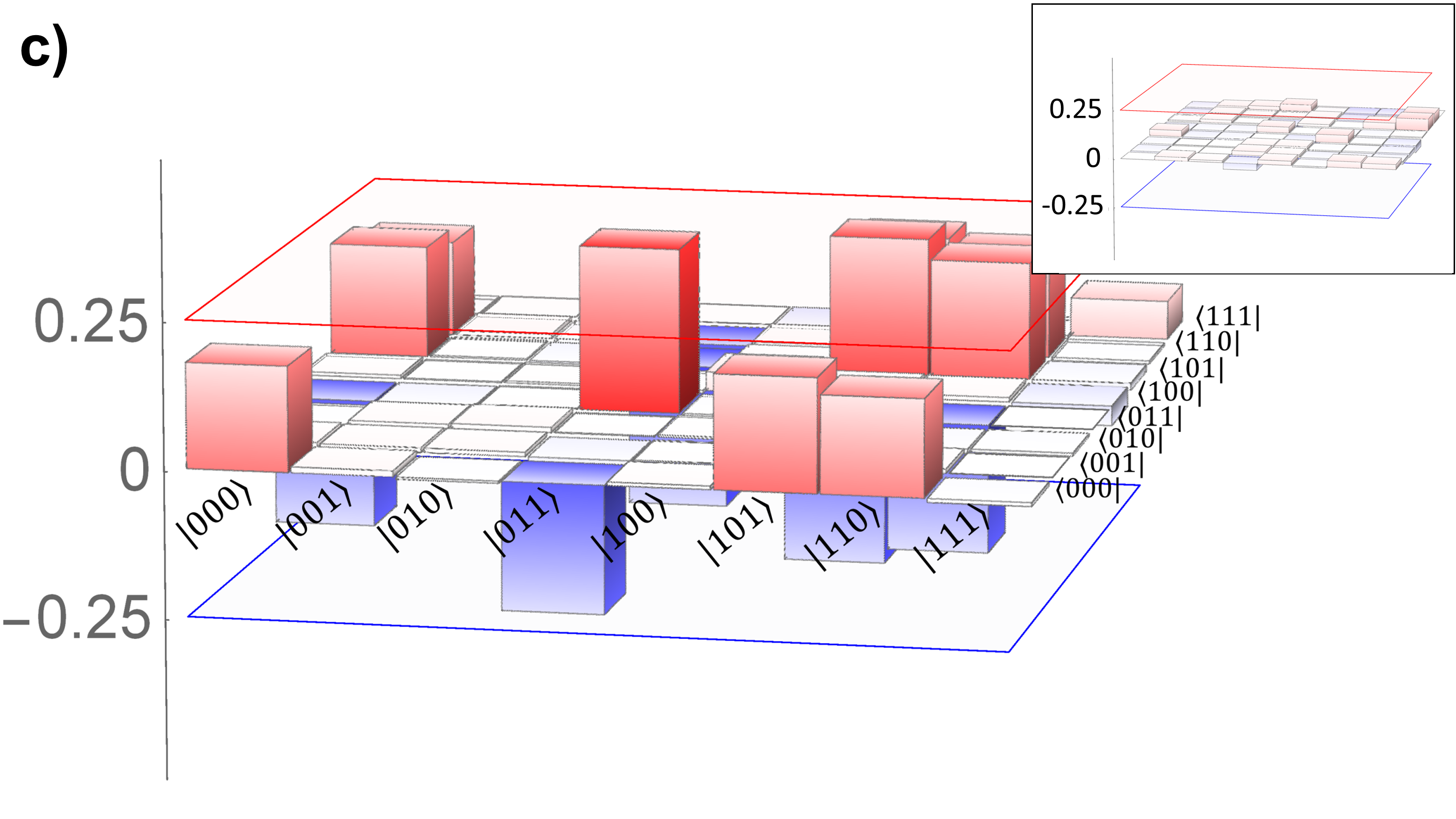}
\includegraphics[width=0.9 \columnwidth, keepaspectratio=true]{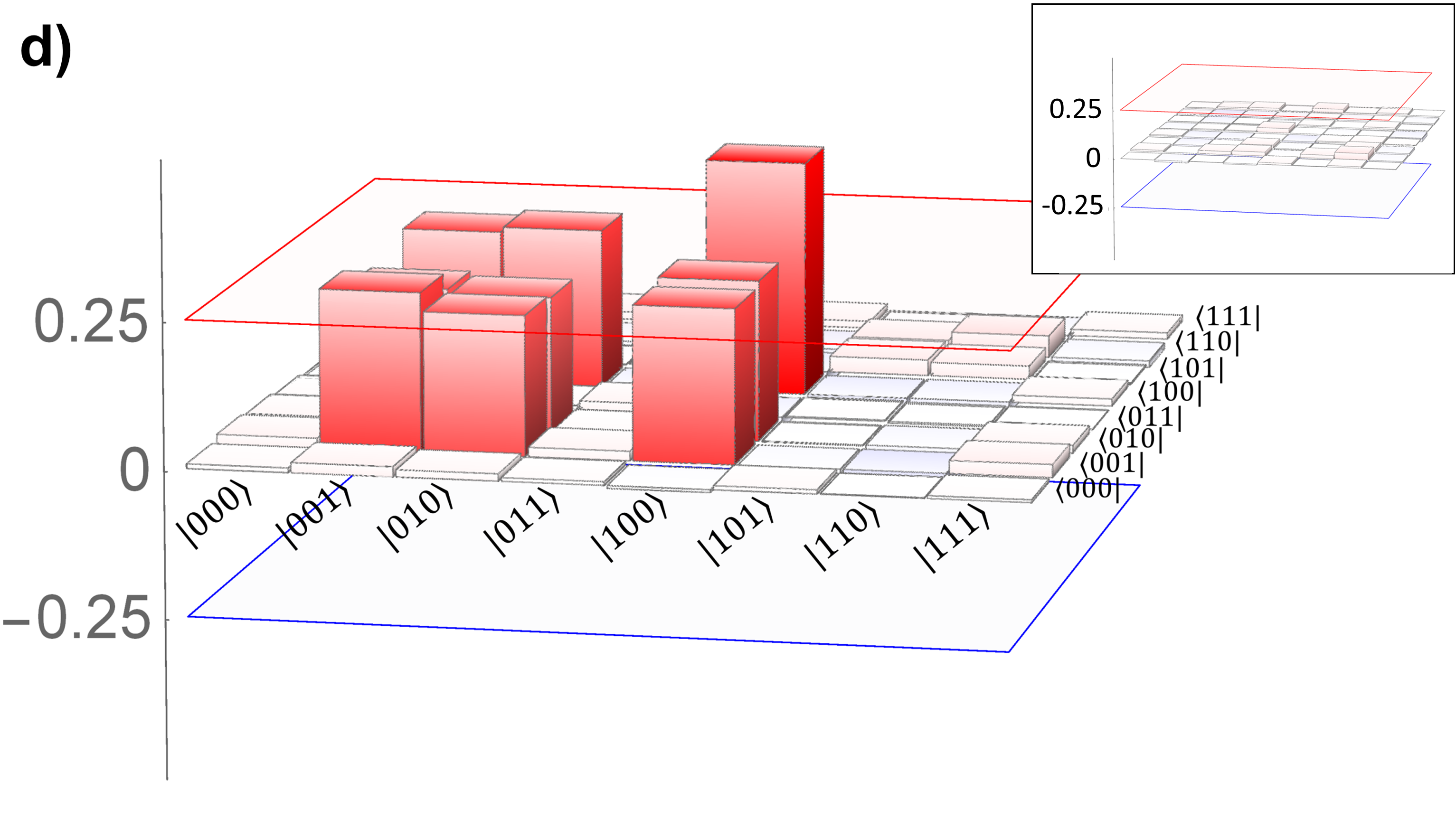}
\caption{Experimental \red{apparatus} for the production of two-photon three-qubit GHZ (a) and W (b) states and the reconstructed density matrices. For both experiments, we start with a two-photon polarisation entangled state coming from the source in~\cite{Ciampini17}. A third qubit is added by manipulating the path of one of the photons inside a displaced Sagnac interferometer. For the GHZ arrangement (a), the path is correlated to the polarisation by using two HWPs. For the W arrangement, polarisation-dependent loss are introduced by nesting a second interferometer, and blocking the $\ket{H}_2\ket{c}_3$. Next c), d) show the  experimental density matrices for the GHZ state and W states respectively, reconstructed by means of a maximum likelihood procedure~\cite{White99}. The full images show the real parts, while the insets report the imaginary part. The unbalance in the populations originated in unwanted polarisation sensitivity of the beam splitter while loss of coherence can be attributed to limited visibility in the Sagnac interferometers. These effects led to measured fidelities $F \equiv \bra{\psi} \rho \ket{\psi} =(83.98\pm0.02)\%$ for the GHZ state, and $F=(81.62\pm0.04)\%$ for the W state.}
\label{figure2}
\end{figure*}

\red{As an example, we consider several three qubit states: namely the GHZ state given by $\ket{\Psi_{\rm{GHZ}}}=\left(|0\rangle |0\rangle |0\rangle + |1\rangle |1\rangle |1\rangle\right)/\sqrt{2}$ and the W state represented as $\ket{\Psi_{\rm{W}}}=\left(|0\rangle |0\rangle |1\rangle + |0\rangle |1\rangle |0\rangle+ |1\rangle |0\rangle |0\rangle\right)/\sqrt{3}$. These `iconic' states can not be transformed into each other by means of local unitary operations, and thus represent different entanglement classes.  In Figure (\ref{idealwigner}) we plot the equal angle Wigner function for both the GHZ and W states. Two immediate observations can be made. First, both show negative region (blue area) clearly indicating the quantum nature of those states; classical states, indeed, can not possess negative regions. This is however not surprising as qubits are uniquely quantum by definition with no classical analog. Further it is important to mention that just because a Wigner function is completely positive, the state may still exhibit some form of non-classicality. Second, the advantage of the Wigner visualisation don't stop at helping in recognising similarities between states. Indeed, in this configuration a procedure known as "Quantum fingerprinting" can be further applied in order to get additional useful information such as entanglement properties and quantum state recognition.}

\section{Generating GHZ and W states}\label{Experiment}

\red{We now turn our attention to simple examples} of how the Wigner function can be used in a real experimental environment. \red{Here, we} produce three-qubit states starting from the two-photon polarization entanglement source demonstrated in~\cite{sorgente}, complemented with a path-encoded qubit~\cite{Ciampini17}. \red{Our} source consists of a 1.5-mm barium borate (BBO) crystal, which, by means of a double-passage arrangement of the pump \red{at 355nm, can produce degenerate} polarisation-entangled photon pairs at 710 nm. The state has the form $\frac{1}{\sqrt{2}}\left(\ket{H}_1\ket{H}_2+\ket{V}_1\ket{V}_2\right)$, where the subscripts 1 and 2 refer to the two polarisation qubits, \red{with $H$ ($V$) indicating} the horizontal (vertical) polarization. The third qubit is added by manipulating the spatial mode of one of the photons \red{(see Fig.~\ref{figure2})} depending on the target state. \red{This represents} the third logical qubit in our state, while still being associate to a physical property of the second photon.

\red{Now,} GHZ states are generated using the interferometer \red{depecited} in Fig.~\ref{figure2}a. Here photon $2$ from the source is sent through a displaced Sagnac interferometer that adds an extra degree of freedom in the clockwise or counterclockwise direction \red{of} the loop. A half-waveplate (HWP) on the anti-clockwise $\ket{a}_3$ path rotates the polarization from $H$ to $V$ and viceversa, while a second HWP on the clockwise path $\ket{c}_3$ introduces a $\pi$ phase shift. \red{Our final state can be written as}:
\begin{align}
\label{TheExpGHZState}
\ket{\psi_{\rm{GHZ}}}=&\frac{1}{2}\left(\ket{H}_1\ket{H}_2\ket{c}_3-\ket{V}_1\ket{V}_2\ket{c}_3 \right. \nonumber \\
&+ \left.\ket{H}_1\ket{V}_2\ket{a}_3+\ket{V}_1\ket{H}_2\ket{a}_3\right).
\end{align}
which is a cluster-form GHZ, locally equivalent to the one defined above

The \red{approach for generating the} W state adopts a similar strategy \red{to the GHZ state}, though the setup is more involved.  Starting with the same two-photon state, photon 2 is now injected in two nested displaced Sagac loops, where the path in the first loop serves as the third qubit as before, while the role of the second loop is to introduce a polarization-dependent loss, as shown in Fig.~\ref{figure2}b. The \red{circuit} generates the state:
\begin{multline}
\label{TheExpWState}
\ket{\psi_{\rm W}}=\frac{1}{\sqrt{3}}\big(\ket{H}_1\ket{H}_2\ket{c}_3 +\ket{H}_1\ket{V}_2\ket{a}_3+\ket{V}_1\ket{H}_2\ket{a}_3\big)
\end{multline}

\red{Concerning our state measurement including our detection setups, spatial} filtering is implemented by means of single-mode fibres \red{with} 5-nm spectral filters adopted. Measurements are conducted \red{using a} standard tomography \red{approach where for the polarisation degree of freedom we measure the projectors associated with the} Stokes parameters while, for the spatial mode, this corresponds to measuring the two outputs of the Sagnac interferometers with different relative phases between the clockwise and anticlockwise paths. 
For both states, we have collected a set of informationally complete measurements that allow us to reconstruct the expressions for the experimental density matrices $\rho_{\rm{GHZ}}$, and $\rho_{\rm{W}}$, respectively, as the output of a maximum likelihood routine. The experimental matrices are shown in Fig.~\ref{figure2} c),d) \red{and at a} glance shows that our experimental states suffer from unbalance in the populations, and from a lack of purity. However, it is hard  \red{to easily determine by visual inspection the class of the state generated or} how the correlations among the qubits will be affected. 

\section{Experimental Sliced Wigner functions}

\begin{figure}[htb]
\begin{center}
{\includegraphics[width=0.9\columnwidth, keepaspectratio=true]{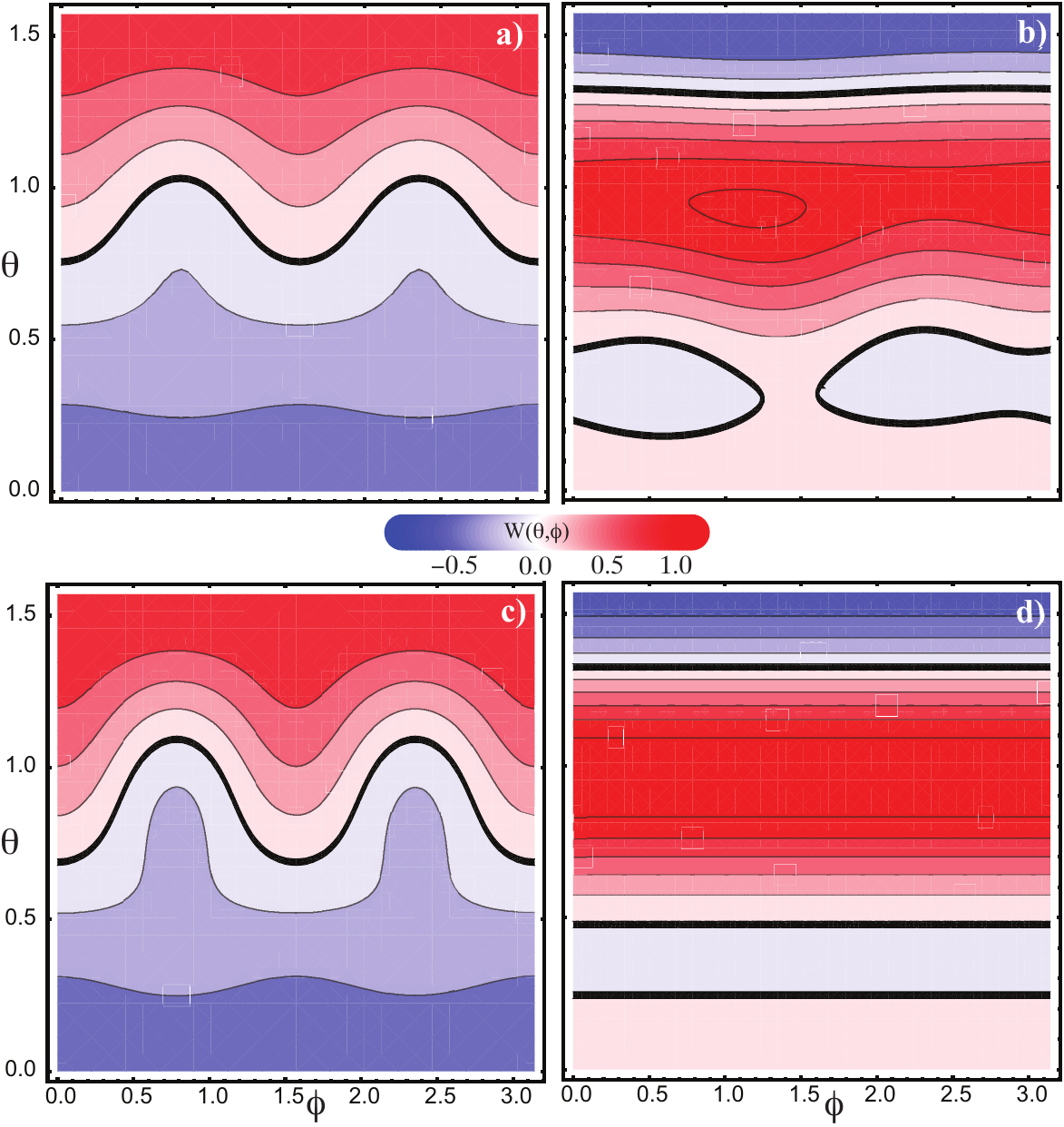}}
\end{center}
\caption{Plot of the magnitude of the equal angle Wigner function reconstructed from experimental data of the cluster GHZ (a) and W (b) states versus $\theta$ and $\phi$. Here the red indicates where the Wigner function is positive while blue where it is negative. Further white represent the boundary where the Wigner function is near zero with the black line(s) indicated where the Wigner function is exactly zero.Next the Wigner functions for the ideal  cluster GHZ (c) and W (d) states are shown for comparison. The colour scaling is the same as in Figure 2. }
\label{TheGHZandWAexp} 
\end{figure}

\red{The Wigner function will provide a natural way to explore these two experimentally realized states. It is important to mention again that our GHZ we generated is not the typical one of the form $\ket{\Psi_{\rm{GHZ}}}$  but instead the cluster state variant (local operations obviously transforms one to the other). In Figure (\ref{TheGHZandWAexp}) we plot both the reconstructed cluster GHZ and W states along side their ideal states for comparison.}
\red{It is immediately clear that in this equal angle slice, the Wigner function of the typical GHZ state and the cluster GHZ state appear different even though the cluster GHZ state is related to usual GHZ state by $\ket{\psi_{\rm{GHZ}}} = P_1 P_2 P_3^\dagger H_1 H_2 H_3\ket{\Psi_{\rm{GHZ}}}$, where $P_i$ and $H_i$ are the usual phase and Hadamard gates acting on the $i$-th qubit. Second, our agreement between the ideal and reconstructed Wigner function is quite remarkable even though we have mentioned issues associated with experimental imperfections and noise. There are of course differences but the overall similarities are clear. }  

\red{The effect of imperfections is generally that of smoothing features and disrupting symmetries. An example of the first instance is visible for GHZ states: the oscillations in Fig.~\ref{TheGHZandWAexp}a exhibit a pattern which is less pronounced that in its theoretical counterpart, Fig.~\ref{TheGHZandWAexp}c. For W states, instead, the expected symmetry in $\phi$ is manifestly broken, as the comparison between Fig.~\ref{TheGHZandWAexp}b and d show. It is interesting to remark how imperfections tend to make the experimental GHZ more symmetric in $\phi$, introducing a hint of a W-like behaviour, while, conversely, the experimental W state starts displaying some oscillations proper to the GHZ state.}

\red{It is interesting to explore the effect of noise on the negative nature of these Wigner functions. It is immediately clear that the depth of our negative regions has decreased due to noise and imperfections. The interesting question is what happens as the amount of noise increases. The depth of the negative region in both the GHZ and W states keep decreasing until we reach a critical point at which it completely disappears in the sliced section of this Wigner function. This critical point occurs near the separability point of our states and so is a useful tool to explore the `quantumness' of these discrete variable states. We must remember that however that the separable system may still be `quantum' due to it qubit substructure and so show negativity. }

\red{It is now worth while to return to this concept of fingerprinting of quantum states outlining how it could work. Figure \ref{TheGHZandWAexp} clearly show that with this equal angle slide that the states look quite different in that representation. They can thus be used to potentially distinguish them and potentially establish which state they arose from. It is useful at this stage to explore a number of different equal angle slices given by $\theta,\phi$ for both of cluster GHZ and W states; these are obtained by applying a local unitary to one of the qubit. The results are shown in appendix (A) and  clearly indicate that several of the GHZ and W slides look the same. However there are slices where the patterns are quite different which can be used to distinguish them. This highlights they may be useful in fingerprinting those state. Finally we need to consider that the traditional and cluster GHZ states looks quite different in the original slice we presented.}

\section{Discussion and Conclusion.}

\red{To summarize, we have shown how the sliced qubit based Wigner function is a useful visual tool for distinguishing different types of quantum states.  Further the present of negative regions in the sliced Wigner function is a clear signature of `quantumness'. While such `quantumness'  is usual in qubit based system, our slicing technique may provide a simple way to distinguish mixed entangled states from their separable counterparts.
}

\red{While providing a conclusive methodology for distinguishing high-dimensional quantum states based on our wigner representation will be left for future works, it is worth mentioning the strength of this approach. In Fig. 3 and 4 we showed that given a specific equal-angle slicing we are able to disintguish between GHZ and W form, including their cluster variant. In higher dimension, the classification of the states become rapidly more complex due to the richness of their inner quantum correlations.  What we propose is to take "snapshots"of different equal-slicing angles (as shown for example for the simple cases considered in this article in the figure in the appendix), providing a set of fingerprints of the quantum state. The characteristics of each of those snapshots overall provides with the complete information about the state, thus allowing to identify it even without previous knowledge on the state itself. Even more, we envision that image recognition algorithms will be able to provide an automatic machine-learning enhanced classification of those states given the topological properties of the fingerprinting. } 
\acknowledgments
We would like to thank R. Rundle for productive discussions. 
T.T. and K.N. acknowledge that this work was supported in part by JSPS KAKENHI (C) Grant Number JP17K05569.
M.A.C. acknowledges support from QUCHIP-Quantum Simulation on a Photonic Chip, FETPROACT-3-2014, Grant agreement no: 641039.


\appendix
\section{Fingerprints}\label{appendixa}
\begin{figure}[htb]
\begin{center}
\subfigure[]{\includegraphics[width=.32\columnwidth, keepaspectratio=true]{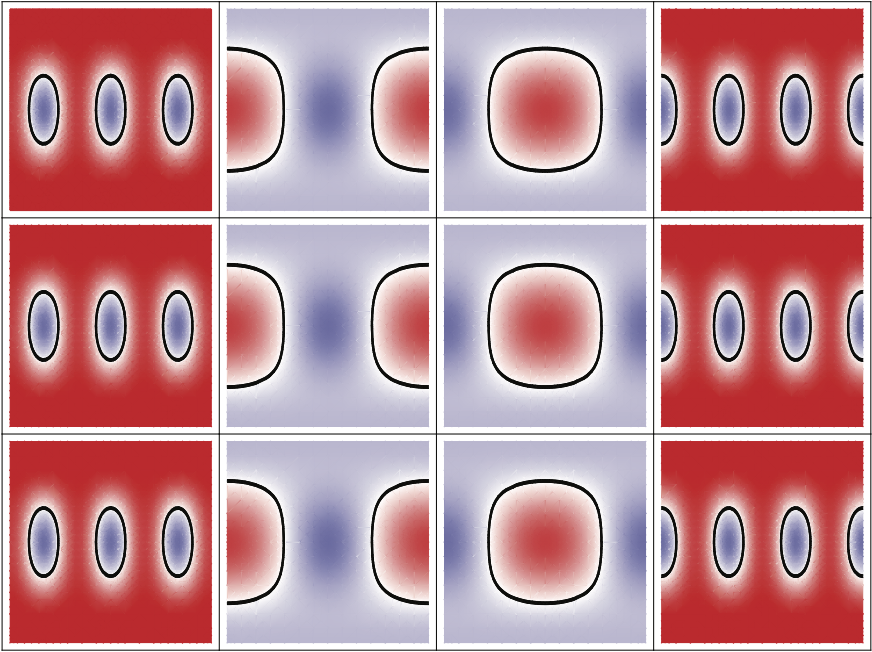}}
\subfigure[]{\includegraphics[width=.32\columnwidth, keepaspectratio=true]{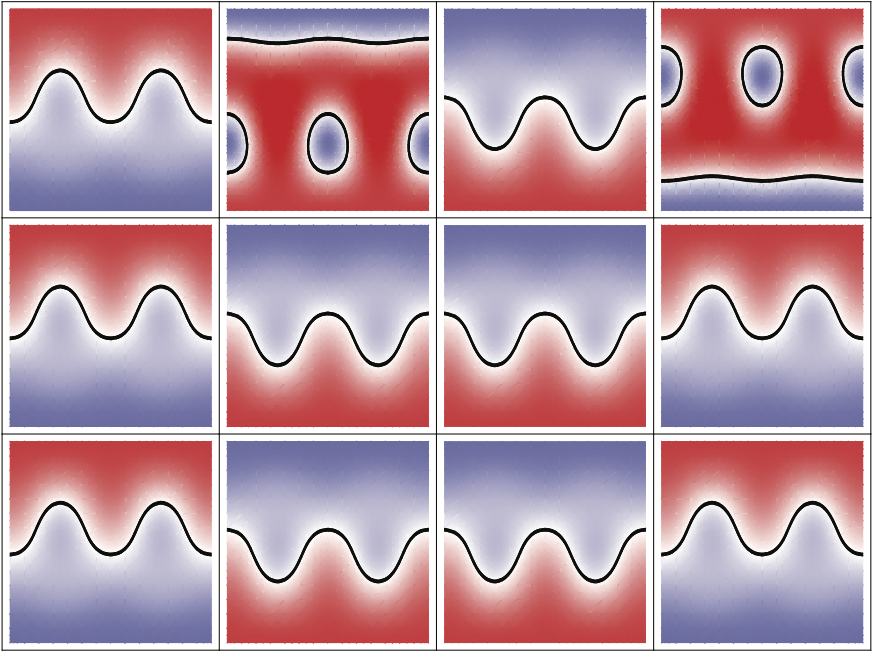}}
\subfigure[]{\includegraphics[width=.32\columnwidth, keepaspectratio=true]{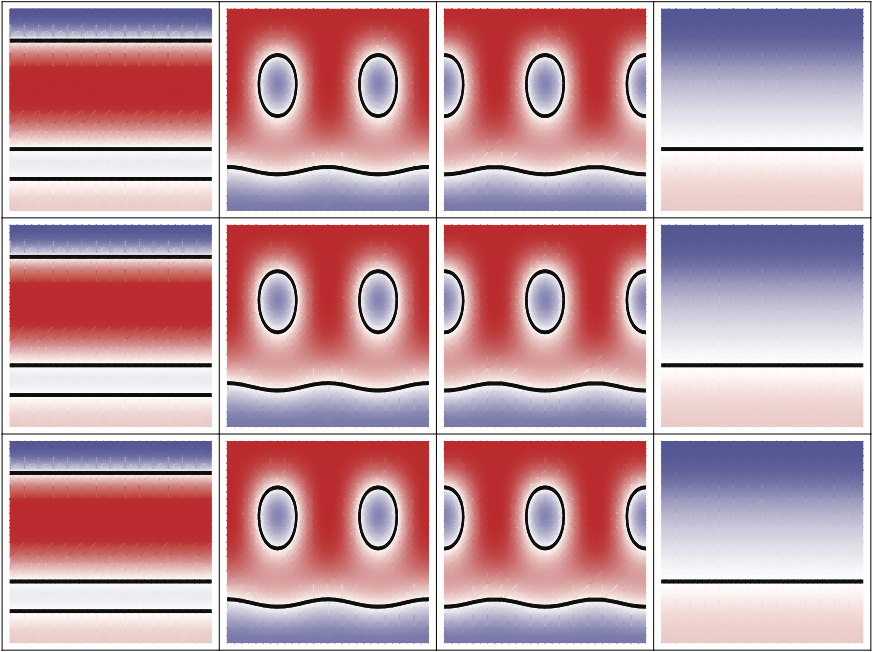}}\\
\end{center}
\caption{Equal angle Wigner function slices of a) the GHZ, b) cluster GHZ and c) W states against $\theta$ (y-axis) and $\phi$ (x-axis). Each subfigure is arranged as a 4x3 set of images where either a $I_i$, $X_i$, $Y_i$ or $Z_i$ rotation is applied to the $i$-th qubit before the equal angle Wigner function is calculated. Each rows corresponds to a different rotated qubit $i$ ranging from 1 to 3. For symmetric states under permutation of the qubits (GHZ and W states), the images in the different row are identical as expected.}
\label{figappendix} 
\end{figure}

\red{In this appendix we graphically present in Fig \ref{figappendix}) a number of equal angle slices for both  GHZ, cluster GHZ and W states. It is clear that many of these images are unique to the state they came from. }

\bibliographystyle{apsrev4-1}
\bibliography{Main-Bib}


\end{document}